# Mechanically Controlled Quantum Interference in Individual π-stacked Dimers

Riccardo Frisenda[1], Vera Jansen[1], Ferdinand C. Grozema[2,*], Herre S.J. van der Zant[1,*] and Nicolas Renaud[2]

[1] *Kavli Institute of Nanoscience, Delft University of Technology, Lorentzweg 1, 2628 CJ Delft, The Netherlands.*
[2] *Department of Chemical Engineering, Delft University of Technology, Julianalaan 136, 2628 BL Delft, The Netherlands.*

*E-mail: h.s.j.vanderzant@tudelft.nl; f.c.grozema@tudelft.nl*

## Abstract

Recent observations of destructive quantum interference in single-molecule junctions confirm the role played by quantum effects in the electronic conductance properties of molecular systems. We show here that the destructive interference can be turned ON or OFF within the same molecular system by mechanically controlling its conformation. Using a combination of ab-initio calculations and single-molecule conductance measurements, we demonstrate the existence of a quasi-periodic destructive quantum interference pattern along the breaking traces of π-π stacked molecular dimers. The detection of these interferences, which are due to opposite signs of the intermolecular electronic couplings, was only made possible by a combination of wavelet transform and higher-order statistical analysis of single-breaking traces. The results demonstrate that it is possible to control the molecular conductance over a few orders of magnitudes and with a sub-angstrom resolution by exploiting the subtle structure-property relationship of π-π stack dimers. These large conductance changes may be beneficial for the design of single-molecule electronic components that exploit the intrinsic quantum effects occurring at the molecular scale.

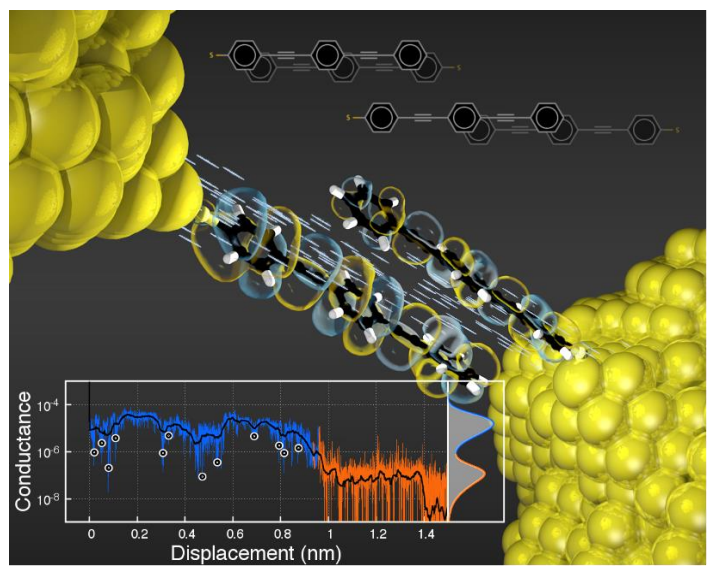

**Introduction**

Charge transport across non-covalently bonded molecular fragments is a crucial process in chemistry, biology and material science. For example, the propagation of charge along π-stacked assemblies plays an important role in the oxidative damage and repair of DNA,[1,2] participates in the early stages of charge separation in the photosynthesis apparatus[3,4] and controls the efficiency of organic materials for electronic applications.[5,6] The efficiency of charge transport through π-π stacked molecules is controlled by the electronic coupling between neighboring molecules and therefore depends in an intricate way on the distance and respective orientation of the π-systems.[7,8,9] While the impact of such molecular packing on charge mobility is well understood for organic crystals[5], only a few experiments have probed the structure-property relationship of π-stacked dimers at the single molecule level[10-14]. A better understanding of coherent charge transport through π-stacked molecules could therefore be beneficial to the engineering of new materials and may also help understanding fundamental charge transfer processes in chemistry.

In this contribution we study charge transport through π-π stacked molecules at the single-molecule level with the mechanically controlled break junction technique[15] (MCBJ). In the experiment, a π-π stacked dimer bridges two lithographically defined metallic electrodes[16,17] that are progressively moved away from each other with sub-angstrom control. The motion of the electrodes can be used to manipulate the dimer conformation by slowly separating the two monomers. The current passing through the molecular structure is continuously measured during the dimer separation by the application of a small bias voltage across the junction. The experiments reveal the presence of pronounced conductance drops when the two π-systems slide over each other. A higher-order statistical analysis of the experimental data, guided by electronic structure calculations, leads to the conclusion that these drops result from destructive quantum interference effects.[18-20] The results therefore show that destructive quantum interference effects can be turned ON and OFF within the same system by mechanically controlling its conformation. The large ON/OFF conductance ratio can be used to design single-molecule electronic[12,20-23] or thermoelectric[11] components that exploit the quantum effects occurring at the molecular scale.

**Theory**

The molecules studied in this work are shown in Fig. 1a. They consist of an oligo-phenylene-ethynylene (OPE3) π-conjugated molecule, with either one (S-OPE3) or two (S-OPE3-S) thiol anchoring groups. Previous studies have demonstrated that single S-OPE3-S molecules reproducibly form molecular junctions in which a single molecule bridges the two electrodes.[24] In contrast, S-OPE3 only contains a single anchoring group and therefore two molecules forming a π-stacked dimer are required to create a mechanically stable electronic connection between the two electrodes[10,13] as illustrated in Fig. 1b.

The electronic transport properties of S-OPE3 dimers and S-OPE3-S were calculated at the density functional level of theory (DFT) using the Landauer formalism[25] in the wide-band limit approximation.[26] The main results of these calculations are summarized in Fig. 1c. In the case of S-OPE3-S molecules, ab-initio calculations at the optimized junction geometry yield a conductance of $2\times10^{-5}$ $G_0$, where $G_0$=77 μS is the quantum of conductance. In the case of a π-π stacked S-OPE3 dimer, the conductance was calculated as a function of displacement of the two molecules with respect to each other (see Figure 1b). As shown in Figure 1c the resulting zero-bias conductance strongly depends on the stacking geometry of the molecules, with very sharp quasi-periodic conductance drops over several orders of magnitude and separated by 0.21 – 0.25 nm. These conductance drops can be traced back to destructive quantum interference effects that occur exactly at the Fermi energy of the electrodes for specific dimer conformations (see Fig. S1 and S2). Additionally, transport calculations combined with molecular dynamics simulations clearly demonstrate that these pronounced modulations of the conductance due to quantum interference are still statistically detectable, even at room temperature (Fig. S10). These modulations of the conductance are preserved when shifting the band alignment between the electrodes and the molecules as illustrated by the blue shaded area in Fig. 1c. However this effect can be significantly blurred if one of the two molecules interacts with both electrodes as represented in Fig. S28.

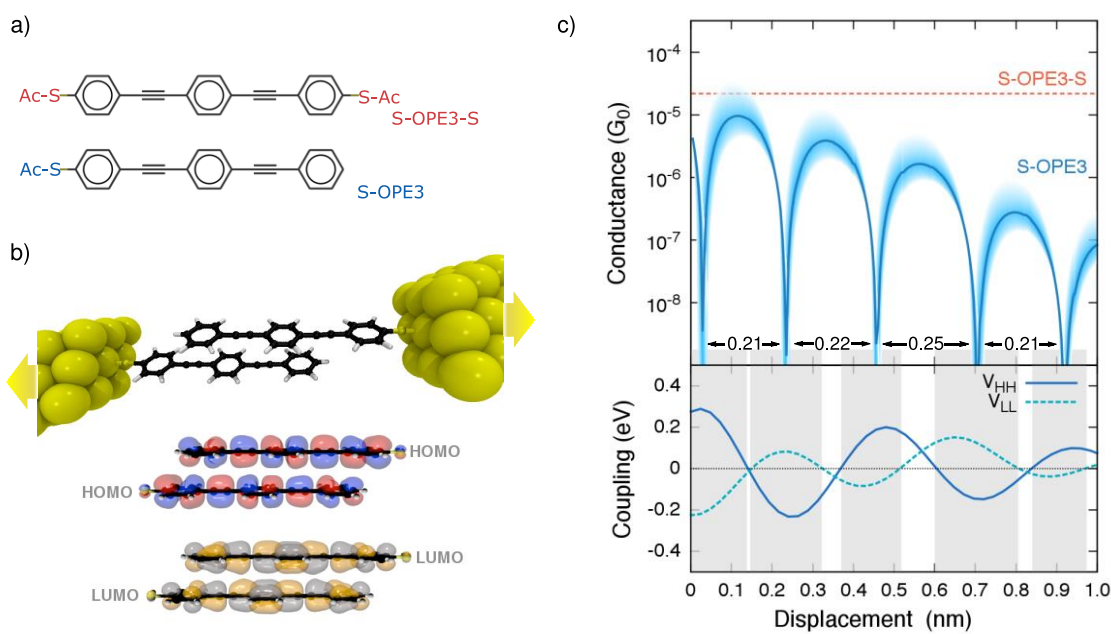

**Figure 1: Theoretical calculations.** a) Chemical structure of the two molecules studied in this article. b) Representation of the two electrodes in the break junction connected by a π-stacked dimer. Fragment molecular orbitals of both dimers are shown. c) Top panel: Conductance of S-OPE3 in the wide-band limit approximation calculated at the Fermi energy (plain blue line). The shaded area around it corresponds to a variation of the Fermi energy of $\pm0.5$ eV. The horizontal dashed line marks the conductance of S-OPE3-S. Bottom panel: Intermolecular electronic couplings between the HOMOs (plain line) and LUMOs (dashed line) fragment molecular orbitals of S-OPE3.

Our calculations show that the couplings between the frontier molecular orbitals play a crucial role in these quantum interference effects. A representation of these orbitals, i.e. the HOMO and LUMO, on each monomer is shown in Fig. 1b. The variations of the electronic coupling between the HOMO of one molecule and the HOMO of the other one, as well as the LUMO-LUMO coupling are shown in Figure 1c. Both vary strongly with the stacking geometry of the dimer, taking alternatingly positive and negative values. This alternation can be understood from the structure of the molecular orbitals (Fig. 1b). Depending on the relative displacement of the two molecules, the overlap of their HOMOs can either be positive, leading to a negative coupling, or negative, which leads to a positive coupling. This argument also holds for the two LUMOs.

By comparing the two graphs in Fig. 1c, one can see that the sharp drops observed in the conductance of the S-OPE3 dimer occur within the regions where the HOMO-HOMO and LUMO-LUMO electronic couplings have opposite sign (grey shaded area in Fig. 1c). As seen in Fig, S1b, a destructive interference pattern is always present somewhere in the gap of the junction when the couplings have opposite signs. The study of a model four-level quantum system reported in Fig. S5, confirms the direct link between the opposite sign of the intermolecular electronic couplings and the presence of destructive quantum interference. Since the values of these couplings are dictated by the overlap between the molecular orbitals, the observation of a quasi-periodic destructive interference pattern constitutes a unique way to probe the structure and phase of the fragment orbitals. As a consequence this effect is very general and can be observed with any dimers of π-conjugated systems, albeit with different periodicities (see e.g. Fig. S11). This effect can even be generalized to more complex systems such as the trimers and tetramers represented in Figs. S25-S27.

**Break-Junction Experiments**

Measurements of the conductance through S-OPE3 dimers and S-OPE3-S molecules were performed using the mechanically controlled break-junction (MCBJ) technique.[27] In this technique, a nanometer-sized gap in a lithographically defined gold nanowire is repeatedly opened and closed in the presence of molecules that can be trapped between the electrodes. The continuous measurement of the molecular conductance while displacing the electrodes results in a so-called breaking trace.

We have measured 1878 breaking traces for S-OPE3 and 1000 for S-OPE3-S. The breaking speed is chosen to be 3-10 pm/s, about two orders of magnitude slower than in usual experiments to obtain detailed information of the conductance traces. Representative traces are shown in Figure 2a. For both S-OPE3 and S-OPE3-S stable conductance plateaus are observed after opening of the nanogap, indicating the formation of a single-molecule junction. The breaking traces of S-OPE3 are generally longer than those of S-OPE3-S and exhibit a more complex structure with multiple conductance plateaus and pronounced conductance drops. These drops show a reduction of the conductance by two orders of magnitude on displacement of the electrodes by a few hundreds of picometer (inset of Fig. 2a). As seen in Fig. 2a. the breaking traces end abruptly when the contact between the two electrodes is disrupted either due to the detachment of one molecule from the electrode or due to the rupture of the π-stacking between two monomers. At this point the motion of the electrodes is reversed until the gap in the gold wire is completely closed and a new opening/closing cycle starts.

To obtain a representative picture of the conductance properties of both molecules, we have constructed two-dimensional conductance-displacement histograms from the full data sets of breaking traces. These histograms, shown in Fig 2b, indicate how often a certain combination of conductance and displacement occurs. Clear differences are observed in the histograms for the two molecules. For S-OPE3-S, a single region of high counts is observed that extends to ≈0.5 nm near $10^{-4}$ $G_0$. This conductance value is comparable with previous studies of the S-OPE3-S single-molecule conductance[24] and with our theoretical calculations. In contrast, the conductance histogram for S-OPE3 exhibits two characteristic regions of high probability, one with a high-conductance state around $10^{-5}$ $G_0$ and a lower one around $10^{-7}$ $G_0$. These two regions extend for about 1.0 nm and 1.7 nm respectively. A few traces even extend up to 3.5 nm, much longer than the length of a single S-OPE3 molecule (1.8 nm). A similar structure, showing two preferential values of the dimer conductance, was also observed in our combined molecular dynamics/transport calculations for S-OPE3 dimers (Fig. S10). Additionally, the measured values of the zero-bias conductance are in good agreement with our DFT simulations that includes explicit gold clusters to model the electrodes (Fig. S2 and S4b).

The marked differences between the two histograms constitute a strong indication that fundamentally different molecular structures are probed in presence of the S-OPE3 and S-OPE3-S. In addition, the long traces observed for S-OPE3, extending up to about twice the length of single molecules, strongly suggest that π-π stacked dimers are responsible for the conductance in these experiments. The double-plateau structure observed both experimentally and theoretically indicates the presence of at least two preferential inter-molecular conformations of the two molecules, each with a specific conductance value.

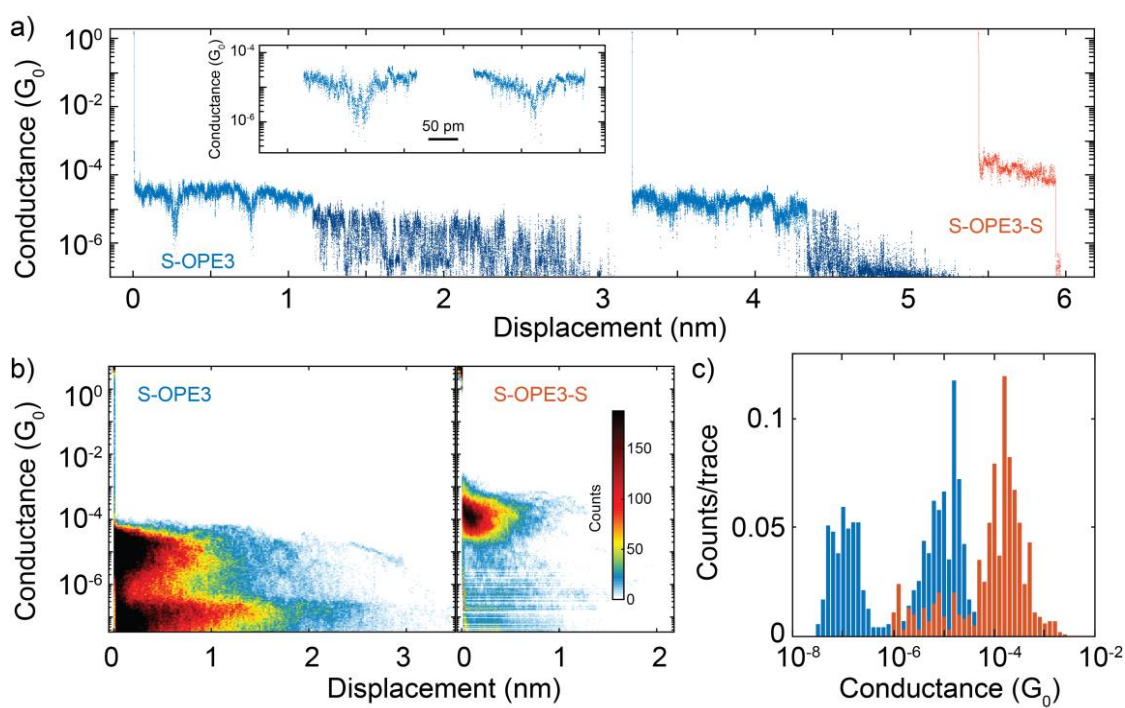

**Figure 2: Conductance analysis of S-OPE3 π-stacked dimers.** a) Examples of breaking traces measured in presence of S-OPE3 (blue) and S-OPE3-S (red). The bias voltage is 100 mV and the electrode-separation speed is 3 pm/s in the first case and 10 pm/s in the second case. Inset: zoom-in on the drops observed in the mono-thiol molecule. b) Two-dimensional conductance-displacement histogram of S-OPE3 and S-OPE3-S built from 1878 and 1000 consecutive traces respectively. A small number of counts at low conductance values is also observed for S-OPE3-S; these may be due to dimers formed by these molecules, albeit with a low probability. c) Conductance histograms built from the mean conductance of each plateau found from a single-trace analysis.

**Single-Trace Statistical Analysis**

While the conductance histograms reported in Fig. 2b give statistical information across the full data set, they provide little insight in conductance variations along individual breaking traces.[28] To extract this information from our experimental results we have performed an extensive statistical analysis of single traces. We have first identified the different conductance plateaus along the breaking traces of both molecules and computed the mean value and standard deviation of their conductance level as well as their total length. Histograms of these mean conductance values are reported in Fig. 2c. Scatter plots of the plateaus length and standard deviations of their conductance level are plotted versus their mean conductance values in Fig. 3a and 3b respectively.

This analysis confirms that the breaking traces of S-OPE3-S mainly contain short plateaus with an average length of 0.29 nm and a mean value of the conductance of $2\times10^{-4}$ $G_0$. The length of 0.29 nm indicates that in contrast to junctions probed in fast-breaking experiments, junctions with slow breaking speeds are not extended over their full length but that premature molecule detachment from the electrodes occurs. In contrast, the plateaus of S-OPE3 are clearly separated into two groups. One group presents a conductance value of $1\times10^{-5}$ $G_0$ and a standard deviation of 0.41 decades while the other shows a lower mean value of $2\times10^{-7}$ $G_0$ with a larger standard deviation of 0.65 decades. The average lengths of the plateaus in the two groups are 0.47 nm and 0.62 nm respectively. These results again point to the existence of two π-π stacked conformations of S-OPE3 with different binding energies. This is consistent with our theoretical calculations that show two local minima of the dimer binding energy separated by about 0.7 nm (Fig. S6). The slight difference between experimental results and theoretical predictions of the plateau length may come from the diffusion of the molecules at the surface of the electrodes. The large standard deviation observed in the second conductance plateau can be explained by the small binding energy of weakly overlapping dimers.

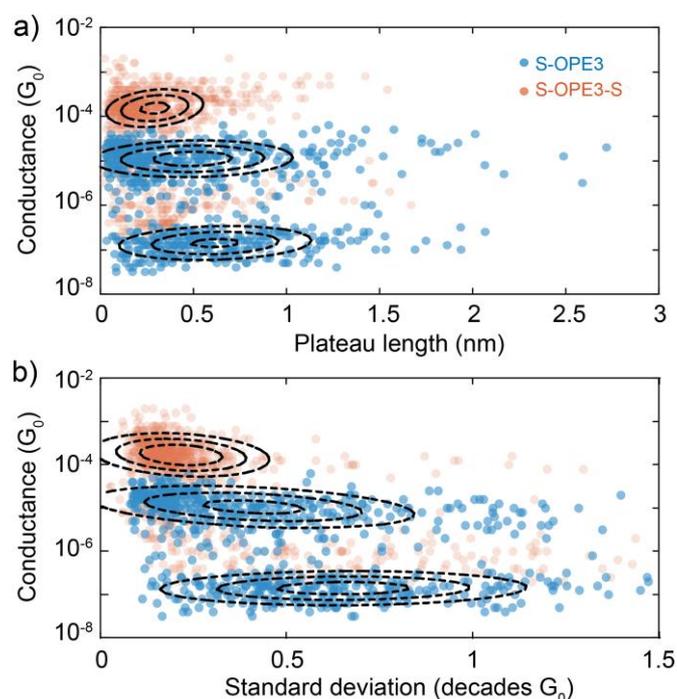

**Figure 3: Statistical analysis of the conductance plateaus.** Scatter plot of the plateau length (a) and of the standard deviation (b) versus the conductance from the analysis of all the plateaus measured for S-OPE3 (blue) and S-OPE3-S (red). Each dot represent a single-conductance plateau extracted from the breaking traces. The dashed lines represent the contour level of bivariate normal distributions. One single bivariate normal peak is sufficient to fit the S-OPE3-S distribution while two peaks are needed for S-OPE3.

As mentioned above, sharp conductance drops highlighted in Fig. 2a are observed for the breaking traces of S-OPE3. These conductance drops occur almost exclusively in the presence of S-OPE3 and hence appear to be characteristic of π-stacked dimers. To determine whether these drops are due to quantum interference effects, the distances between consecutive drops were extracted from our experimental results and compared with our theoretical prediction of 0.21-0.25 nm. Given the low standard deviation of their conductance level, the high-conductance plateaus of 132 S-OPE3 traces were considered (Fig. S20). A higher-order statistical analysis of single breaking traces (Fig. S21), adapted from detection techniques developed for biological signals,[29] was implemented to automatically detect the sharp conductance drops (see supplementary section VI for more details). The position and depth of the drops were then used to compute the distribution of drop-to-drop distances in the experimental data. Representative examples of this analysis are shown in Fig. 4a for four different breaking traces of S-OPE3. The same analysis was also performed on 259 breaking traces of S-OPE3-S for comparison (Fig. S24).

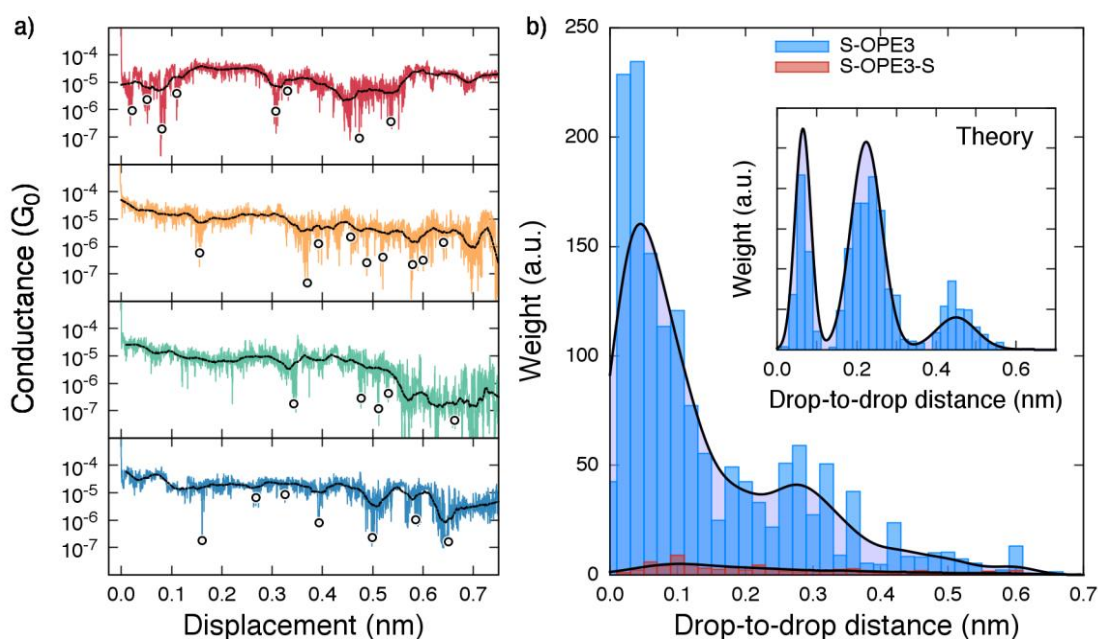

**Figure 4: Statistical analysis of the conductance drops observed in single traces.** a) Automatic detection of conductance drops on a few conduction traces of S-OPE3. b) Histograms of the weighted drop-to-drop distance obtained from the statistical analysis, performed with 131 traces of S-OPE3 and 259 traces of S-OPE3-S. The black lines show a kernel density estimation fitting of the distribution. The inset shows the histogram obtained for the theoretical traces of S-OPE3.

The histograms of the drop-to-drop distance obtained for both molecules are shown in Fig. 4b. As seen in this figure our statistical analysis clearly demonstrates that the conductance drops are predominantly found in S-OPE3. Although some drops were also found along the breaking traces of S-OPE3-S, they are much less pronounced than those found for S-OPE3 molecules (see Fig. S24). The statistical analysis of single traces further reveals that the drop-to-drop distance distribution for the S-OPE3 breaking traces is not random. A kernel density estimation fitting of the distribution, shown as a full black line in Fig. 4b, leads to two peaks located at 0.05 nm and 0.27 nm respectively. As demonstrated in Supplementary section IV-D, the first peak can be attributed to an intrinsic limitation of the detection algorithm. This is confirmed by the appearance of a similar peak in the distribution of the drop-to-drop distances obtained for theoretical traces in the presence of noise as shown in the inset of Fig. 4b. Interestingly, the position of the second peak in the experimental distribution is close to the theoretical prediction of 0.21-0.25 nm. Considering that several factors may influence the drop-to-drop distance, we find the agreement between the theoretical prediction and the experimental detection satisfactory. These factors include uncertainties related to the theoretical treatment (level of theory, presence of gold clusters, position of Fermi energy) but also physical issues such as the possibility that a bound molecule moves to a different position on the surface of the gold electrode during the recording of the breaking trace. The latter can lead to an apparent increase or decrease of the distance between consecutive conductance-drops.

We finally note that a wide range of mechanisms can also induce a significant reduction of the conductance. This is for example the case of an increase of the π-π stacking distance between the two monomers or the deviation of one molecule from the axis of the junction (see Fig. S7). However, as argued in the supplementary section II, the conductance drops induced by these mechanisms are expected to occur at random places, in contrast with the distribution shown in Fig. 4b. The statistical analysis developed in this article consequently allows distinguishing between these undesirable conductance drops and those induced by quantum interference effects.

**Conclusion**

We have presented a combined theoretical and experimental study of electronic transport across π-π stacked dimers. The theoretical analysis reveals the presence of quasi-periodic conductance drops along breaking traces of π-stacked dimers. These drops are the results of destructive quantum interference that occurs when the electronic couplings between the frontier molecular orbitals of the two monomers exhibit opposite signs. As a consequence, the periodicity of the drops is dictated by the structure and phase of the frontier molecular orbitals and does not correspond to the distance between benzene rings. Using a mechanically controlled break-junction experiment we have measured the electronic conductance of molecular π-π stacked dimers while slowly pulling the stacked molecules apart. Large conductance drops were recorded in the breaking traces. A higher-order statistical analysis of the experimental results shows a striking resemblance with the theoretical predictions. The results thus demonstrate the possibility to control molecular conductance over a few orders of magnitudes and with sub-angstrom resolution by exploiting the subtle structure/property relationship of π-stacked systems. This effect is not limited to the specific molecular structure studied here and a similar behavior is expected for a large class of molecules. The future study of mechanically controlled quantum interference in π-stack systems might be facilitated by exploiting specific molecule-molecule interactions that increase the binding energy between them or the use of organic cages to partially encapsulate the monomers.

**Methods**

**Theoretical Calculations** Calculations of the electronic structure were performed at the DFT level of theory using the Amsterdam Density Functional (ADF) Package[30]. A double-zeta basis set with polarization was used in combination with the M06-2X functional. The transport calculations in the wide band limit were done using a home-built program. The electronic coupling matrix elements were calculated within the ADF package by using the orbitals of the two monomers as basis and by extracting the off-diagonal elements of the Fock matrix[31]. The MD+Transport calculations were performed with a modified version of the TransPull program[32]. Explicit frozen gold clusters were considered in these calculations to simulate the bulk electrodes. A detailed discussion of the theoretical methods used here can be found in supplementary section I.

**MCBJ Experiments** In the MCBJ experiment, a lithographically fabricated nano-scaled gold wire was patterned on a flexible substrate. Electrodes are first characterized without molecules to ensure that they are clean and that they are well aligned (in which case a clear peak at 1 $G_0$ is observed indicative of atomically sharp electrodes). After that, a droplet of the solution containing the molecules (1 mM in dicloromethane) is deposited on the electrodes; after evaporation of the solvent, measurements are performed in air at room temperature. More details can be found in section V-B of the supporting information and a detailed description of this technique has been given previously[33]. We investigate the low-bias conductance of the two OPE3 molecules at a bias of 100 mV using a low pulling speed of the electrodes of 3 pm/s in the case of S-OPE3 and 10 pm/s for S-OPE3-S. This slow pulling speed in combination with the stability of the MCBJ technique enabled us to acquire sufficient amounts of data points per breaking trace to facilitate further statistical analysis. All experiments were performed at room temperature. Similar conductance experiments with S-OPE3c were carried at lower temperatures (T=77 K and T=195 K), but π-stacked dimer configurations were only observed at room temperature (see supplementary section V-C). We measured in total 1878 conductance breaking traces in presence of S-OPE3 and 1000 with S-OPE3-S. The molecular junction formation yield was 25% in the S-OPE3 case and 90% for S-OPE3-S, corresponding respectively to 469 molecular junctions in the first case and 901 in the second case.

**Analysis of the plateau length** To analyze the plateaus lengths and standard deviations we first compute the conductance histogram of each individual breaking trace. We then locate the different peaks of this distribution and their half-width and half-maximum (HWHM). After denoising and decimation of the breaking trace we locate the regions where the conductance signal is within two HWHM around the conductance peak center. We then identify the different plateaus in the original trace and extract their length and the standard deviation of the conductance. We checked for the algorithm consistency and robustness and we found that most of the traces are well analyzed; in the few cases of error we correct manually the entries.

**Statistical Analysis of Conductance Drops** We have carefully selected 132 traces among the full data set of S-OPE3 breaking traces. As illustrated in Fig. S20, these traces were selected based on their total length, number of plateaus and signal-to-noise ratio. The identification of conductance drops along these individual traces was performed via the detection algorithm outlined in Fig. S21. This algorithm is based on the stationary wavelet transform of the baseline corrected breaking traces. The wavelet coefficients were then subjected to a 2-stage Kurtosis denoising[29] to isolate the drops from the noise regions. Burst of the cumulative denoised trace were then detected using an infinite hidden Markov chain[34] for the identification of unique conductance drops. The weighted distribution of drop-to-drop distance was then computed via the position and depth of each individual drop. The continuous probability function of the drop-to-drop distance distributions was obtained via a kernel density estimation fitting. Our detection method was proven to be robust with respect of various parameters involved in our algorithm as illustrated in Fig. S22.

**Electronic Supplementary Information** (ESI) available at the address
http://www.nature.com/nchem/journal/v8/n12/abs/nchem.2588.html

**References**

(1) Núñez, M. E.; Hall, D. B.; Barton, J. K.: Long-range oxidative damage to DNA: Effects of distance and sequence. *Chemistry & Biology* **1999**, *6*, 85-97.

(2) Merino, E. J.; Boal, A. K.; Barton, J. K.: Biological contexts for DNA charge transport chemistry. *Current Opinion in Chemical Biology* **2008**, *12*, 229-237.

(3) Brettel, K.; Leibl, W.: Electron transfer in photosystem I. *Biochimica et Biophysica Acta (BBA) - Bioenergetics* **2001**, *1507*, 100-114.

(4) Wasielewski, M. R.: Photoinduced electron transfer in supramolecular systems for artificial photosynthesis. *Chemical Reviews* **1992**, *92*, 435-461.

(5) Coropceanu, V.; Cornil, J.; da Silva Filho, D. A.; Olivier, Y.; Silbey, R.; Brédas, J.-L.: Charge Transport in Organic Semiconductors. *Chemical Reviews* **2007**, *107*, 926-952.

(6) Sirringhaus, H.; Brown, P. J.; Friend, R. H.; Nielsen, M. M.; Bechgaard, K.; Langeveld-Voss, B. M. W.; Spiering, A. J. H.; Janssen, R. A. J.; Meijer, E. W.; Herwig, P.; de Leeuw, D. M.: Two-dimensional charge transport in self-organized, high-mobility conjugated polymers. *Nature* **1999**, *401*, 685-688.

(7) Yi, Y.; Coropceanu, V.; Bredas, J.-L.: A comparative theoretical study of exciton-dissociation and charge-recombination processes in oligothiophene/fullerene and oligothiophene/perylenediimide complexes for organic solar cells. *Journal of Materials Chemistry* **2011**, *21*, 1479-1486.

(8) Solomon, G. C.; Herrmann, C.; Vura-Weis, J.; Wasielewski, M. R.; Ratner, M. A.: The Chameleonic Nature of Electron Transport through π-Stacked Systems. *Journal of the American Chemical Society* **2010**, *132*, 7887-7889.

(9) Delgado, M. C. R.; Kim, E.-G.; Filho, D. A. d. S.; Bredas, J.-L.: Tuning the Charge-Transport Parameters of Perylene Diimide Single Crystals via End and/or Core Functionalization: A Density Functional Theory Investigation. *Journal of the American Chemical Society* **2010**, *132*, 3375-3387.

(10) Wu, S.; Gonzalez, M. T.; Huber, R.; Grunder, S.; Mayor, M.; Schonenberger, C.; Calame, M.: Molecular junctions based on aromatic coupling. *Nat Nano* **2008**, *3*, 569-574.

(11) González, M. T.; Leary, E.; García, R.; Verma, P.; Herranz, M. Á.; Rubio-Bollinger, G.; Martín, N.; Agraït, N.: Break-Junction Experiments on Acetyl-Protected Conjugated Dithiols under Different Environmental Conditions. *The Journal of Physical Chemistry C* **2011**, *115*, 17973-17978.

(12) Fujii, S.; Tada, T.; Komoto, Y.; Osuga, T.; Murase, T.; Fujita, M.; Kiguchi, M.: Rectifying Electron-Transport Properties through Stacks of Aromatic Molecules Inserted into a Self-Assembled Cage. *Journal of the American Chemical Society* **2015**, *137*, 5939-5947.

(13) Martín, S.; Grace, I.; Bryce, M. R.; Wang, C.; Jitchati, R.; Batsanov, A. S.; Higgins, S. J.; Lambert, C. J.; Nichols, R. J.: Identifying Diversity in Nanoscale Electrical Break Junctions. *Journal of the American Chemical Society* **2010**, *132*, 9157-9164.

(14) Batra, A.; Kladnik, G.; Vázquez, H.; Meisner, J. S.; Floreano, L.; Nuckolls, C.; Cvetko, D.; Morgante, A.; Venkataraman, L.: Quantifying through-space charge transfer dynamics in π-coupled molecular systems. *Nat Commun* **2012**, *3*, 1086.

(15) Tao, N. J.: Electron transport in molecular junctions. *Nat Nano* **2006**, *1*, 173-181.

(16) Li, Q.; Solomon, G. C.: Exploring coherent transport through [small pi]-stacked systems for molecular electronic devices. *Faraday Discussions* **2014**, *174*, 21-35.

(17) Li-Li, L.; Xiu-Neng, S.; Yi, L.; Chuan-Kui, W.: Formation and electronic transport properties of bimolecular junctions based on aromatic coupling. *Journal of Physics: Condensed Matter* **2010**, *22*, 325102.

(18) Solomon, G. C.; Andrews, D. Q.; Goldsmith, R. H.; Hansen, T.; Wasielewski, M. R.; Van Duyne, R. P.; Ratner, M. A.: Quantum Interference in Acyclic Systems: Conductance of Cross-Conjugated Molecules. *Journal of the American Chemical Society* **2008**, *130*, 17301-17308.

(19) Guedon, C. M.; Valkenier, H.; Markussen, T.; Thygesen, K. S.; Hummelen, J. C.; van der Molen, S. J.: Observation of quantum interference in molecular charge transport. *Nat Nano* **2012**, *7*, 305-309.

(20) Sautet, P.; Joachim, C.: Electronic interference produced by a benzene embedded in a polyacetylene chain. *Chemical Physics Letters* **1988**, *153*, 511-516.

(21) Stafford, C. A. C., D. M.; Mazumdar, S.: The quantum interference effect transistor. *Nanotechnology* **2007**, *18*, 42014.

(22) Baer, R.; Neuhauser, D.: Phase Coherent Electronics: A Molecular Switch Based on Quantum Interference. *Journal of the American Chemical Society* **2002**, *124*, 4200-4201.

(23) Perrin, M.; Frisenda, R.; Koole, M.; Seldenthuis, S.; Valkenier, H.; C., H. J.; Renaud, N.; Grozema, F. C.; Thijssen, J. M.; Dulic, D.; vd Zant, H. S. J.: Large negative differential conductance in single-molecule break junctions. *Nat. Nano.* **2014** *9*, 830-834.

(24) Frisenda, R.; Tarkuç, S.; Galán, E.; Perrin, M. L.; Eelkema, R.; Grozema, F. C.; van der Zant, H. S. J.: Electrical properties and mechanical stability of anchoring groups for single-molecule electronics. *Beilstein Journal of Nanotechnology* **2015**, *6*, 1558-1567.

(25) Cuevas, J. C.; Scheer, E.: *Molecular Electronics: Anintroduction to Theory and Experiment*; World Scientific, 2010.

(26) Verzijl, C. J. O.; Seldenthuis, J. S.; Thijssen, J. M.: Applicability of the wide-band limit in DFT-based molecular transport calculations. *The Journal of Chemical Physics* **2013**, *138*, 094102.

(27) He, J.; Sankey, O.; Lee, M.; Tao, N.; Li, X.; Lindsay, S.: Measuring single molecule conductance with break junctions. *Faraday Discussions* **2006**, *131*, 145-154.

(28) Frisenda, R.; Perrin, M. L.; Valkenier, H.; Hummelen, J. C.; van der Zant, H. S. J.: Statistical analysis of single-molecule breaking traces. *Physica status solidi (b)* **2013**, *250*, 2431-2436.

(29) Brychta, R. J.; Shiavi, R.; Robertson, D.; Diedrich, A.: Spike detection in human muscle sympathetic nerve activity using the kurtosis of stationary wavelet transform coefficients. *Journal of Neuroscience Methods* **2007**, *160*, 359-367.

(30) Theoretical Chemistry, V. U., Amsterdam, The Netherlands, http://www.scm.com: ADF 2014. 2014.

(31) Senthikumar, K.; Grozema, F. C.; Bickelhaupt, F. M.; Siebbeles, L. D. A.: Charge transport in columnar stacked triphenylenes: Effects of conformational fluctuations on charge transfer integrals and site energies. *J. Chem. Phys* **2003**, *119*, 9809.

(32) Hutcheson, J.; Franco, I.; Renaud, N.; Carigano, M.; Ratner, M. A.; Schatz, G. C.: TransPull. 2011.

(33) Martin, C. A.; Smit, R. H. M.; Egmond, R. v.; van der Zant, H. S. J.; van Ruitenbeek, J. M.: A versatile low-temperature setup for the electrical characterization of single-molecule junctions. *Review of Scientific Instruments* **2011**, *82*, 053907.

(34) Kleinberg, J.: Bursty and hierarchical structure in streams. In *Proceedings of the eighth ACM SIGKDD international conference on Knowledge discovery and data mining*; ACM: Edmonton, Alberta, Canada, 2002; pp 91-101.

**Acknowledgement**

The research leading to these results has received funding from the European Research Council FP7 ERC Grant Agreement no. 240299 (Mols@Mols) and Horizon 2020 ERC Grant Agreement no. 648433.

**Author contributions**

N.R. and F.C.G. performed the electronic transport calculations and the molecular dynamics simulations. R.F., V.J. and H.S.J.Z. performed the break-junction experiments. R.F. designed and implemented the analysis of the plateaus. N.R. designed and implemented the higher-order statistical analysis of conductance drops. All authors wrote the manuscript.

**Competing financial interests**

The authors declare no competing financial interests.